\documentclass[a4paper,11pt]{article}
\usepackage{jcappub} 
\usepackage{lineno}
\usepackage{booktabs}
\usepackage{multirow}

\usepackage{xcolor}

\usepackage{lscape} 

\newcommand{\ha}[1]{#1} 
\newcommand{\fede}[1]{#1} 
\newcommand{\fabian}[1]{#1} 

\newcommand{\hess}[1]{H.E.S.S.}
\newcommand{\mmw}[1]{multi-wavelength}

\title{\boldmath H.E.S.S. {programme} searching for VHE {gamma rays} associated {with} FRBs}

\author[1,2,3]{F.~Aharonian,}
\author[4]{A.Archaryya,}
\author[5]{J.~Aschersleben,}
\author[6,*]{H.~Ashkar, \note[*]{Corresponding authors}}
\author[7,8]{M.~Backes,}
\author[9]{V.~Barbosa~Martins,}
\author[10]{R.~Batzofin,}
\author[11,12]{Y.~Becherini,}
\author[9,13]{D.~Berge,}
\author[2]{K.~Bernl\"ohr,}
% ...
\affiliation[1]{Dublin Institute for Advanced Studies, 31 Fitzwilliam Place, Dublin 2, Ireland}
\affiliation[2]{Max-Planck-Institut f\"ur Kernphysik, Saupfercheckweg 1, 69117 Heidelberg, Germany}
\affiliation[3]{Yerevan State University,  1 Alek Manukyan St, Yerevan 0025, Armenia}
\affiliation[4]{University of Southern Denmark}
\affiliation[5]{Kapteyn Astronomical Institute, University of Groningen, Landleven 12, 9747 AD Groningen, The Netherlands}
\affiliation[6]{Laboratoire Leprince-Ringuet, École Polytechnique, CNRS, Institut Polytechnique de Paris, F-91128 Palaiseau, France}
\affiliation[7]{University of Namibia, Department of Physics, Private Bag 13301, Windhoek 10005, Namibia}
\affiliation[8]{Centre for Space Research, North-West University, Potchefstroom 2520, South Africa}
\affiliation[9]{Deutsches Elektronen-Synchrotron DESY, Platanenallee 6, 15738 Zeuthen, Germany}
\affiliation[10]{Institut f\"ur Physik und Astronomie, Universit\"at Potsdam,  Karl-Liebknecht-Strasse 24/25, D 14476 Potsdam, Germany}
\affiliation[11]{Université Paris Cité, CNRS, Astroparticule et Cosmologie, F-75013 Paris, France}
\affiliation[12]{Department of Physics and Electrical Engineering, Linnaeus University,  351 95 V\"axj\"o, Sweden}
\affiliation[13]{Institut f\"ur Physik, Humboldt-Universit\"at zu Berlin, Newtonstr. 15, D 12489 Berlin, Germany}
% ...

\emailAdd{contact.hess@hess-experiment.eu}

\abstract{
Fast Radio Bursts (FRBs) are highly energetic, extremely short-lived bursts of radio flashes. Despite extensive research, the exact cause of these outbursts remains speculative. The high luminosity, short duration, and high dispersion measure of these events suggest they result from extreme, \ha{high-energy extragalactic} {sources}, {such as} highly magnetized and rapidly spinning neutron stars known as magnetars{.} The number of detected FRBs, including repeating ones, has grown rapidly in recent years. Except for \ha{FRB~20200428D}, and \ha{FRB-like radio burst that is associated {to G}alactic magnetar SGR~1935+2154}, no \mmw{}  {counterpart} to any FRB has been detected yet.  The High Energy Stereoscopic System (H.E.S.S.) telescope has developed a {program} to follow up FRBs {searching for their gamma-ray counterparts, helping to uncover the nature of FRBs and FRB sources.} 
This paper provides an overview of the searches for FRB sources conducted by \hess{}, including follow-up observations and simultaneous \mmw{}  campaigns with radio and X-ray observatories. {Among the FRB sources observed by H.E.S.S., \fede{nine} are localized with redshifts ranging between 0.11 and 0.492 from 2015 to 2022}. No significant very high energy (VHE) emission was detected during these observations. We report constraints on the VHE luminosity ranging from $10^{44}$~erg s$^{-1}$ and $10^{48}$~erg s$^{-1}$, \fede{placing limits on the FRB's region persistent VHE emission and potential FRB afterglow emission across timescales from hours to years.}}

\begin{document}
\maketitle
\flushbottom

\section{Introduction} 

Fast radio bursts (FRBs) are discovered transient phenomena characterized by the sudden appearance of bright and short-duration radio pulses in the sky~\citep{2013Sci...341...53T,2019A&ARv..27....4P,2019ARA&A..57..417C}. These bursts exhibit a broad range of luminosity distances, ranging from \fede{3.6~Mpc~\citep{2021ApJ...910L..18B, 2022Natur.602..585K, 2025NatAs...9..111P} to $\sim$7~Gpc~\citep{2023Sci...382..294R}}, and have short durations, typically spanning from \fede{tens of nanoseconds~\cite{Majid_2021, 2022NatAs...6..393N, 2023NatAs...7.1486S} to milliseconds with few events of duration exceeding a second~\cite{frbcatchime}}. Their luminosities vary between \fede{$10^{38}$~\cite{2022NatAs...6..393N, 2022Natur.602..585K}} and $10^{46}$~erg~s$^{-1}$~\citep{lml+20}.
They were serendipitously discovered with the Parkes Radio Telescope~\cite{Lorimer2007, thornton2013} and are now routinely detected \fede{by various radio facilities, such as CHIME~\cite{chime}}. 
Over a thousand radio bursts from approximately {700} published FRB sources have been reported \citep[see e.g. FRB catalogs][]{frbcat, frbcatchime}.


Some \fede{FRB sources} exhibit recurring bursts. \fede{Among the currently known sample, approximately $\sim 2-3\%$ FRBs has been observed to repeat~\cite{2023ApJ...947...83C}}, leaving the question of whether all FRBs exhibit this {behaviour} open~\citep[e.g.][]{css18}. The distinction between repeating and non-repeating sources and the underlying physical differences between them remain unclear. 
Sources of repeating FRBs have been documented emitting across a wide frequency spectrum, ranging from 110~MHz~\citep{pmb+21} to as high as 8~GHz~\citep{GSP+18}. 
Several sources of repeating FRBs exhibit a regular pattern in their activity cycles. An example is the source of FRB~20180916, as reported in~\citep{2020Natur.582..351C}, and of the initial repeater, FRB~20121102, where a periodicity with a significance of $\sim3\sigma$ was \fede{noted~\citep{2021MNRAS.500..448C, 2020MNRAS.495.3551R}}. More recently, in addition to accounting for the multiple components within individual bursts, a compelling periodicity of 216.8 ms, with a significance of 6.5$\sigma$, was detected in the pulse profile of FRB~20191221A. Such a brief, distinct periodicity strongly supports the idea of a neutron star origin for this event~\citep{2022Natur.607..256C}.
\fede{In order to measure the distance to FRBs,  the dispersion measure (DM), which tracks free electrons along the line of sight, and more recent methods such as repeat pulse observations or interferometric techniques~\citep[e.g.][]{clw+17,doi:10.1126/science.aaw5903} are used.} FRBs are generally thought to originate beyond our 
{Galaxy}. 
Optical observations of well-localized of FRBs have identified 24 FRBs securely associated with host galaxies, with redshift values from $z = 0.03$ to $1.02$~\citep{rbb+22, gfk+23}. \fede{Due to their extragalactic origin, attenuation of their spectra at high energy is expected from photon-photon absorption in the extragalactic background light (EBL)~\cite{2011MNRAS.410.2556D, franceschini_ebl}.} Although most FRBs are extragalactic, \ha{an FRB-like radio burst} has been detected within our {Galaxy} and will be discussed specifically (see \autoref{sec:sgr}).

Numerous theoretical models have been proposed for FRBs \citep{2019PhR...821....1P}, yet the enigma surrounding their origin and emission mechanism remains unsolved (e.g. \citep{2020Natur.587...45Z,2021SCPMA..6449501X,2023RvMP...95c5005Z}).
Compact objects, such as magnetars (strongly magnetized neutron stars) and massive black holes, have gained favor as progenitor models, primarily due to the short-lived nature of FRBs~\cite{2019ARA&A..57..417C}. Some magnetars may produce FRBs {throughout} the interplay of intensely magnetized pulses and the surrounding material, thereby giving rise to synchrotron maser emission {that can extend to TeV {energies}}~\cite{2014MNRAS.442L...9L}. Alternatively, other models propose that repeating FRBs stem from regions near the surface of magnetars, where ultra-relativistic internal shocks and blast waves become associated with flares~\cite{2017ApJ...843L..26B}.

The study of the electromagnetic emission of FRBs across different wavelengths is crucial for unraveling their nature. It allows us to understand source characteristics, explore the surrounding environment, refine emission mechanisms, search for associated phenomena, and explore cosmological and fundamental physics. 
Multiwavelength counterparts to FRBs have remained elusive, with the notable exception of the soft gamma repeater (SGR) SGR 1935+2145 (discussed in detail in \autoref{sec:sgr}{)}, where the detection of an X-ray counterpart is the first evidence that magnetars are linked to sources of repeating FRBs and that non-thermal processes might be behind \fede{FRB emission~\cite{Mereghetti2020, 2021NatAs...5..378L, 2021NatAs...5..401T, Tavani2021}}). 

{The High Energy Stereoscopic System (\hess{}) telescope array, located in the Southern Hemisphere, plays a crucial role in this research by being the only instrument capable of observing transient events in the very high-energy (VHE) gamma-ray band in the southern sky.}
Since 2015, \hess{} has been actively engaged in a comprehensive FRB {programme}, with the aim of investigating potential correlations between these events and VHE gamma-ray emission. {This {programme} employs} two primary strategies: Target of Opportunity (ToO) alerts triggered by radio or X-ray observatories, and coordinated \mmw{}  campaigns.
In the first approach, {the objective is to detect afterglows associated with FRBs and to identify their emission in VHE {gamma rays}.} This involves promptly responding to ToO alerts from radio or X-ray observatories, enabling \hess{} to capture and study any potential gamma-ray emission that may follow the FRB events.
In the second approach, the focus of the {multi-wavelength observations} is on observing known FRB sources, whether repeating or not, to study their host environments {,} and on capturing FRBs simultaneously with radio facilities. Coordinated \mmw{}  campaigns are designed to systematically observe and analyze the activity of these sources. By collecting data across different wavelengths simultaneously, {these campaigns} aim to uncover any recurring patterns or {behaviours} associated with the FRB sources.
{In this paper, after introducing HESS in \autoref{sec:hess}}, {we provide} an overview of the searches for {12} FRB sources conducted by \hess{} since 2015, including follow-up observations (\autoref{sec:followups}) and simultaneous \mmw{}  campaigns (\autoref{sec:campaigns}) with radio and X-ray observatories~\cite{2022A&A...666A.119H}.

\section{The High Energy Stereoscopic System }
\label{sec:hess}
Situated in the Khomas Highland in Namibia at an altitude of 1835~m, \hess{} is an array consisting of five imaging atmospheric Cherenkov telescopes (IACTs)~\cite{hess_aharonian_2006}. Its {main goal} is to detect {VHE} gamma rays spanning energies from a few tens of GeV to 100 TeV. \fede{In comparison, MAGIC~\cite{magic}, with its lower energy threshold (~30 GeV), is particularly suited for detecting soft-spectrum sources, while VERITAS~\cite{veritas}, with a sensitivity similar to H.E.S.S., provides complementary coverage in the Northern Hemisphere.}

The \hess{} array is comprised of two distinct types of telescopes. Telescopes CT1-4 have a field of view (FoV) with a diameter of $5^{\circ}$ and a mirror area of 107~m$^2$. They are arranged in a square formation with side lengths of 120~m. In contrast, a larger telescope, CT5, with a mirror area of 614~m$^2$ and a camera FoV of $3.4^{\circ}$ diameter, is positioned at the {centre} of the array. CT1-4 are dedicated to detecting gamma rays with energies $\gtrsim 100$~GeV, while CT5 is optimized for gamma rays with energies $\gtrsim 30$~GeV{~\cite{ct5}}. 



For transient phenomena occurring on hour timescales, \hess{} exhibits more than 20 times greater energy flux sensitivity in its core energy range above 300 GeV compared to the space-based {\textit{Fermi} Large Area Telescope (LAT)} at 1~GeV energies~\citep[][]{HESS_GRB2021}.
{There are two main instrumental effects that limit \hess{}'s observations: the duty cycle and the minimum detectable energy. \hess{} is operational on nights with clear skies with a moon {phase} below 40\%, which results in a duty cycle ranging between 10\% and 15\% {depending on weather}. A higher duty cycle {allows one} to increase the chances of observations when an alert is received. The minimum detectable energy depends primarily on the mirror size and the zenith angle of observations, with higher zenith angles leading to higher {energy thresholds}. CT5 and its data acquisition system have been fine-tuned for swift repositioning to capture short-lived transients that are expected to rapidly decrease in gamma-ray intensity~\citep{Hofverberg2013,Balzer_2015}. \ha{The lower energy threshold of CT5 is crucial for studying transients of extragalactic origin, as the highest energy emissions from these sources are often absorbed by the EBL (which increases with redshift and energy \cite{2011MNRAS.410.2556D, franceschini_ebl}), leading us to focus on the lower end of the VHE spectrum.}

In the following sections, we describe the two approaches of the {\hess{}} FRB {programme} and the related observations.
{The data described here were processed using the {likelihood reconstruction method} {described} in
\cite{hess_denaurois_2009}.} This method employs standard techniques for gamma-hadron separation and event selection cuts. Background estimation is carried out using the well-established ``reflected background" technique \cite{hess_berge_2007}, and the results are cross-validated through an independent event calibration and reconstruction method \cite{hess_parsons_2014}. 
For the {fluxes or flux limits calculations}, the minimum energy is chosen as the energy where the effective area reaches $10\%$ of its maximum value, {while the maximum energy is chosen when the number of background events $N_{\mathrm{OFF}}$ per energy bin falls under 10.} {The upper limits are calculated using the Rolke method~\cite{hess_rolke_2005} assuming an $E^{-2}$ energy spectrum unless specified otherwise.  \fede{The flux integral upper limit represents the upper bound on the gamma-ray flux above a given energy threshold. It is obtained by integrating the differential flux between the minimum and maximum energies defined above, under the assumption of an $E^{-2}$ spectral shape. } {This assumption comes from the fact that FRBs are extragalactic phenomena and that there is no knowledge of their exact spectral distribution.}


\section{{ToO} follow-up observations}
\label{sec:followups}
{The first strategy in the {\hess{}} FRB {programme} is to conduct automatic follow-up observations of FRBs received as external alerts. {\hess{}}’s ToO alert system receives transient alerts from the Parkes~\cite{2016Natur.530..453K} and the UTMOST  radio telescopes~\cite{utmost} and applies a set of pre-defined criteria to determine when and how to schedule follow-up observations. The criteria are set to maximize science cases {defined} by the H.E.S.S. transients {programme}~\cite{2022A&A...666A.119H}; for FRB alerts, the system is configured to react to both prompt and afterglow triggers.}  {Since 2015, H.E.S.S. has been involved in the follow-ups of three FRBs (see \autoref{ssec:parkes} and \ref{ssec:utmost}) and two magnetars (see \autoref{sec:sgr}).} 


\subsection{Parkes follow-ups: \fede{FRB~20150418A} and \fede{FRB~20150215}}
\label{ssec:parkes}
A first follow-up campaign of FRBs in VHE gamma rays was conducted by \hess{} in response to the detection of \fede{FRB~20150215} and \fede{FRB~20150418A} by {SUPERB (SUrvey for Pulsars and Extragalactic Radio Bursts)~\cite{superb}. {The results of these observations have already been published in \cite{frb20150215} and \cite{frb150418}, respectively.} SUPERB is a pulsar and fast transient survey which utilizes the 64-m Parkes telescope~\cite{2016Natur.530..453K}.}   Since April 2014, the SUPERB project has been dedicated to real-time discovery of FRBs and pulsars~\cite{2016Natur.530..453K}. 

On February 15, 2015, Parkes detected a burst from \fede{FRB~20150215A}, {at a Galactic longitude and latitude of $24.66^{\circ}$, $5.28^{\circ}$ and $25^{\circ}$ away} from the Galactic {Centre} (redshift $z<0.56$), with a DM of $1105.6 \pm 0.8$~pc cm$^{-3}$ and a pulse duration of $2.8^{+1.2}_{-0.5}$~ms, ~\cite{frb20150215}. This burst was promptly detected {in radio} with recorded polarization and \mmw{}  follow-up, including observations with {\hess{}}
Owing to observational constraints, \hess() could only observe the target region one week after the burst. Observations were taken with zenith angles ranging between 54 and 64 degrees under good conditions. The first observation run of 28 minutes commenced on 2015-02-22 at 02:53 UTC, 6.3 days after the Parkes detection and a second observation was conducted on 2015-02-25 starting at 02:49 UTC, 9.3 days after the burst. No gamma-ray emission was detected during both observations.  The resulting \fede{energy flux integral upper limits} at 99\% confidence level (C.L.) were {$\Phi_\gamma(E > 1~\mathrm{TeV}) < 6.38 \times 10^{-14}$~erg~cm$^{-2}$s$^{-1}$}, assuming an $E^{-2}$ energy spectrum~\cite{frb20150215}.

In the same year, {SUPERB detected} a second FRB. The burst \fede{FRB~20150418A} was initially observed at Parkes on April 18, 2015. The fading radio afterglow persisted for up to six days and was associated with an elliptical host galaxy at $z = 0.492 \pm 0.008$ (WISE J071634.59-190039.2)~\cite{frb150418}. 
\hess{} received notification of \fede{FRB~20150418A} from the SUPERB team on 2015-04-18 during the daytime, hindering immediate follow-up observations. The required observation conditions were met later that evening at 17:55 UTC (approximately {14.5~hours} after the FRB), and 1.4 hours of data were recorded until the source set below an elevation of $45^{\circ}$, the typical horizon for observations retaining a relatively low energy threshold.  Since the obtained results were fully consistent with the background expectation, we concluded that no significant gamma-ray afterglow was detected from the direction of \fede{FRB~20150418A}. {Consequently, the 99\% C.L. upper limits using the Feldman Cousin method~\citep{1998PhRvD..57.3873F}} was obtained, assuming a generic $E^{-2}$ energy spectrum for the potential emission, and integrating above the threshold of 350 GeV. This resulted in the integral upper limit on the gamma-ray flux of  $\Phi_\gamma(E > 350 \mathrm{GeV}) < 1.33 \times 10^{-12}~\mathrm{cm^{-2}s^{-1}}$~\cite{frb150418}}.

\subsection{UTMOST follow-up: \fede{FRB~20190806A}}
\label{ssec:utmost}

\begin{figure}[t]
    \centering
    \includegraphics[width=\textwidth]{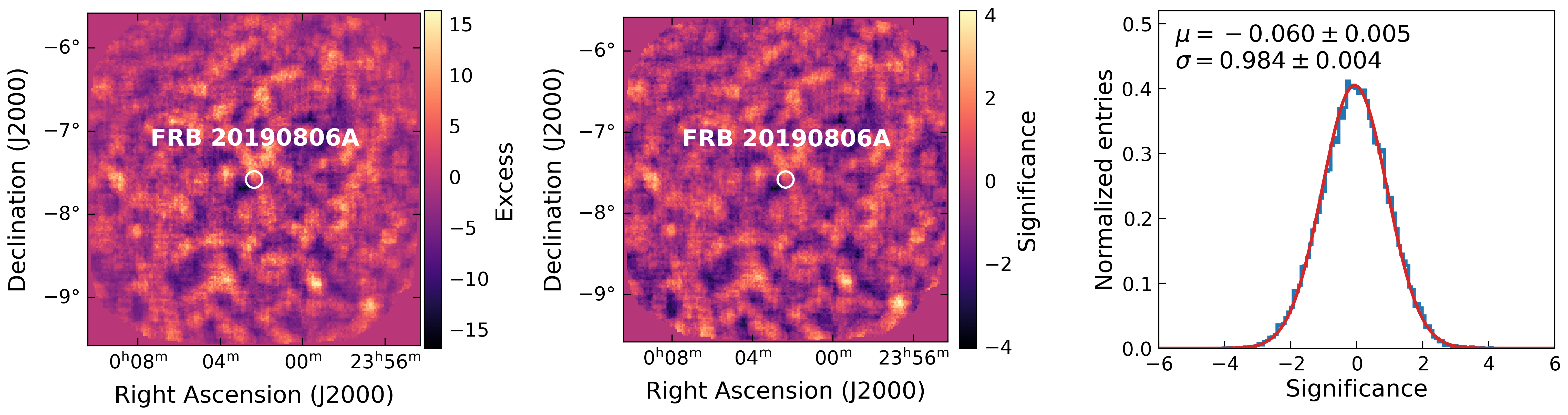}
    \caption{Observations of \fede{FRB~20190806A} triggered by UTMOST. \textbf{Left:} VHE gamma-ray {excess map} around the direction of \fede{FRB~20190806A} illustrated by the event counts exceeding the background. The circle in the {centre} has {a radius of $0.1^{\circ}$} and denotes {the location of the FRB and the search region for a gamma-ray signal.}
    \textbf{Middle:} Significance map of the gamma-ray emission using the formalism proposed by {Li and Ma}~\cite{LiMa} in the region around \fede{FRB~20190806A}. \textbf{Right:} \fede{Significance distribution (blue histogram), normalized by the number of entries and bin width.} The red line and the shown parameters correspond to a Gaussian function fit to the distribution.}
    \label{fig:FRB20190806_HESS_RingMaps}
\end{figure}

UTMOST is the first project to issue {alerts} upon detecting an \fede{FRB~\cite{utmost, 10.1093/mnras/stx638}}. To minimize follow-up latency, \hess{} implemented an automatic response to these alerts. {Since similar portions of the sky are visible to} UTMOST in Australia and \hess{} in Namibia, the goal is to {investigate} the earliest afterglow phase possible. 
For \hess{}, the filtering criteria applied to alerts from UTMOST include the signal-to-noise ratio (S/N), the \textit{importance} and the dispersion measure (DM): 
\begin{itemize}
\item The FRB S/N should be higher than 10; 
\item The \textit{importance} parameter {should be equal to 1. This} indicates if the alert has been vetted by a human;
\item The DM should be  $\mathrm{DM < 1000~pc~cm^{-3}}$. \ha{This is due to the fact that high DM values generally suggest large source distances. Although this is not necessarily the case for every event, we chose to include this filtering to avoid follow-ups of high-redshift FRBs where \hess{} has no realistic chance of detecting VHE gamma-ray emission due to severe EBL absorption.}
\end{itemize}
Moreover, to \ha{conduct FRB follow-ups}, observation conditions, such as darkness and visibility, should be suitable within a few hours {after the alert} (typically up to six hours).

On August 6, 2019, \hess{} received the first alert from UTMOST, \fede{FRB~20190806A}, that passed all filtering criteria described above. FRB~20190816 was detected at 17:07:58.0 UTC with $DM = 388.5~\mathrm{pc~cm^{-3}}$ and the {inferred} redshift is $z<0.32$~\citep{2019ATel12995....1G}. 
{
The \hess{} Transients system~\citep{2022A&A...666A.119H} automatically scheduled follow-up {observations} of \fede{FRB~20190806A} that started 4.5 hours later and lasted 2 hours with the CT1-4 telescopes (CT5 was undergoing a camera {upgrade} at the time and could not participate). {No significant emission of VHE gamma rays was detected (\autoref{fig:FRB20190806_HESS_RingMaps}), but \fede{flux integral upper limits} were derived (\autoref{fig:delays}).} {\autoref{fig:delays} shows the \fede{flux integral upper limits} as a function of the time difference between the gamma-ray observation and the radio burst detection.} This observation was the first successful automatic H.E.S.S. follow-up of an FRB, including alert reception, filtering, scheduling, observations, data transfer, and analysis.}

\begin{figure}
    \centering
    \includegraphics[width=0.9\textwidth]{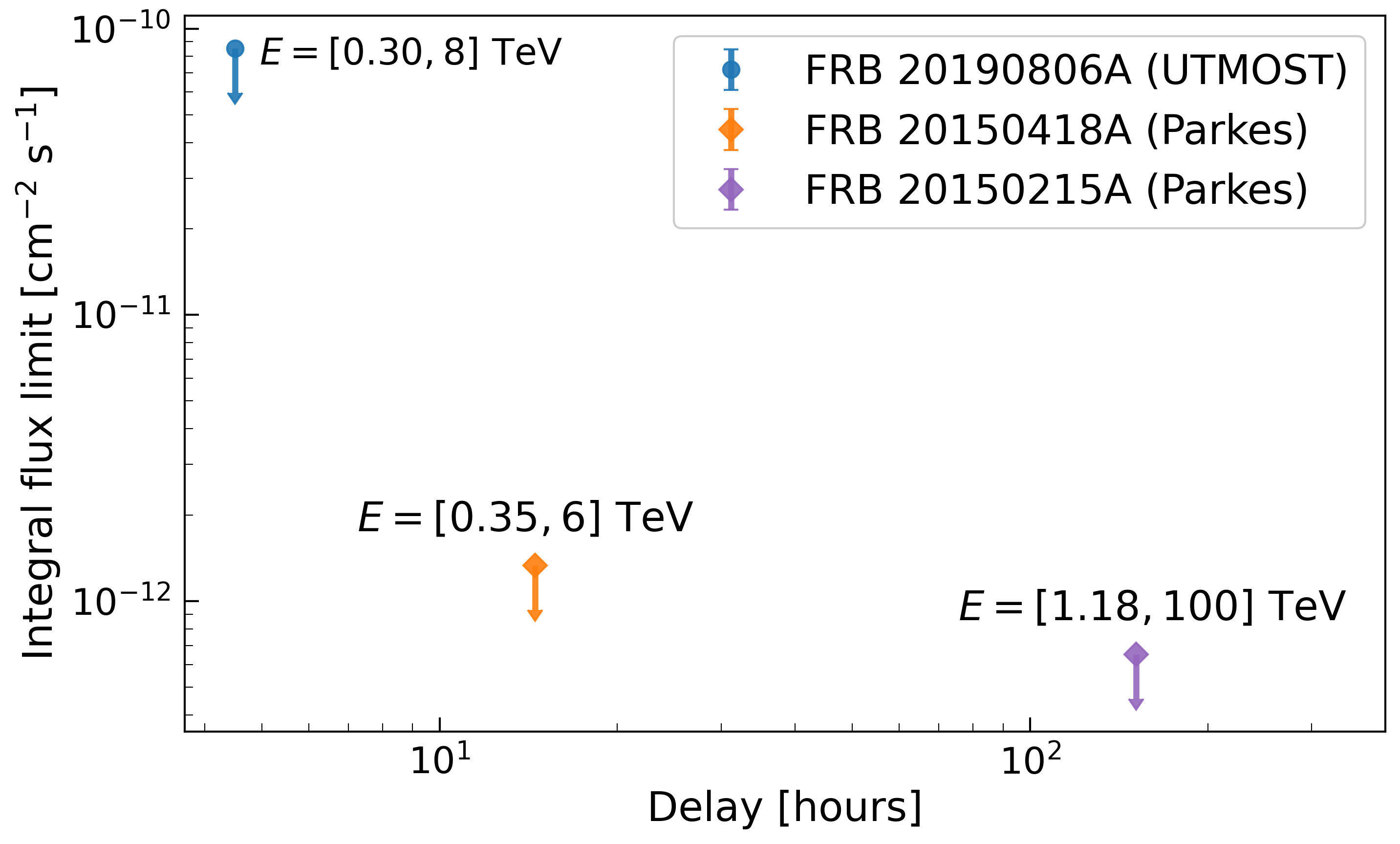}
    \caption{H.E.S.S. observation delays in hours, relative to the radio burst detection of three FRBs by Parkes and UTMOST. \fede{Flux integral upper limits} are shown for each {FRB} within the energy range depicted in the figure. The assumed energy spectrum for these calculations follows an $E^{-2}$ profile. \fede{The timescales used to derive these limits correspond to the H.E.S.S. observation durations for each source, as specified in the main text.}}
    \label{fig:delays}
\end{figure}

\subsection{\fede{Magnetar} observations}
\label{sec:sgr}

The \hess{} search for VHE counterparts of FRBs extends to \fede{magnetars} due to their potential as primary candidate progenitors of FRBs. The goal of magnetar observation campaigns with \hess{} is to search for non-thermal VHE emission from these sources to {have a complete view of} their emission spectrum. 
{This focus arises from a significant event in April 2020, {when} the detection of an FRB coincided with two preceding gamma-ray outburst alerts triggered by the BAT instrument aboard the \textit{Swift} satellite~\citep{2020Natur.587...54C}. Two millisecond-duration radio bursts have been detected from the same direction of SGR J1935+2154{:} the first burst by CHIME~\cite{2020ATel13684....1B} and STARE2~\cite{2020Natur.587...54C}, and the second burst by FAST~\cite{2020ATel13699....1Z}. The timing of the first \fede{FRB-like} radio burst appears to align very well with one of the X-ray bursts seen by AGILE~\cite{Tavani2021}, Konus–Wind~\cite{Ridnaia2021}, INTEGRAL (burst G in \cite{Mereghetti2020}), and Insight-HXMT~\cite{Zhang2020ATel}. \fede{Additionally, between April 2020 and December 2023, six new bursts were detected from SGR~1935+2154 on four different occasions by CHIME~\cite{2023arXiv231016932G}.} This FRB is notable for being the first {Galactic} FRB and the only one detected to date.}

\subsubsection{\fede{SGR~1935+2154: the source of a Galactic FRB-like radio burst}}
Upon the detection of a {Swift X-ray burst~\citep{SWIFT_1_2} }\hess{} observed SGR~1935$+$2154 on 28 April, 2020. Around 2 hours of data were acquired from 01:55 UTC to 03:53 UTC. {A few hours} later an \ha{FRB-like radio burst named} \ha{FRB~20200428D} was detected from the source on April 28 at 14:34:33 UTC. {While the \hess{} observations did not coincide {temporally} with the FRB, they coincided with 4 {other} X-ray bursts detected by Fermi-GBM and INTEGRAL ~\citep{Lin_2020,Mereghetti2020} {from the source.}} No VHE gamma-ray counterparts were detected. An analysis searching for gamma-ray pulses at the millisecond scale did not show any significant pulses coinciding with any X-ray bursts. Details on the analysis and results can be found in~\cite{HESS_SGR1935}. 99\% C.L. {integral} upper limits of ${\Phi_{\gamma}\mathrm{(E > 420\,GeV)} < 1.27\cdot10^{-12}\,\mathrm{cm}^{-2}\,\mathrm {s}^{-1}}$ were set on the {steady} VHE gamma-ray emission from SGR~1935+2154 assuming an $\mathrm{E^{-2}}$ power law spectrum.

\subsubsection{{SGR~1830$-$0645} (ToO 2020-10-10)}
In the scope of the efforts of the \hess{} collaboration to probe VHE gamma-ray emission from \fede{magnetars}, \hess{} observations were triggered on {SGR~1830$-$0645}, a newly discovered magnetar, following the detection of a short burst by Swift~\citep{2020GCN.28594....1P} {on October 10, 2020 at 14:49:24 \fede{UTC}}. Observations started {the same day at 18:13:38 \fede{UTC}} and lasted for 1 hour. Nearby VHE sources were excluded from the analysis. The results do not show any significant VHE gamma-ray signal coming from the {SGR~1830$-$0645} position. 99\% C.L. upper limits of ${\Phi_{\gamma}\mathrm{(E > 284\,GeV)} < 1.78\cdot10^{-12}\,\mathrm{cm}^{-2}\,\mathrm {s}^{-1}}$ were set on the \fede{persistent} VHE gamma-ray emission from SGR~1830-0645 assuming an $\mathrm{E^{-2}}$ power law spectrum.
We show the 99\% C.L. differential upper limits in \autoref{fig:sgr1830_diffULs}. {No continuous VHE sources coincident with \fede{the magnetar's position} {were} detected in the \hess{} {Galactic Plane} survey described in~\cite{2018A&A...612A...1H}.}

\begin{figure}
    \centering    \includegraphics[width=0.9\textwidth]{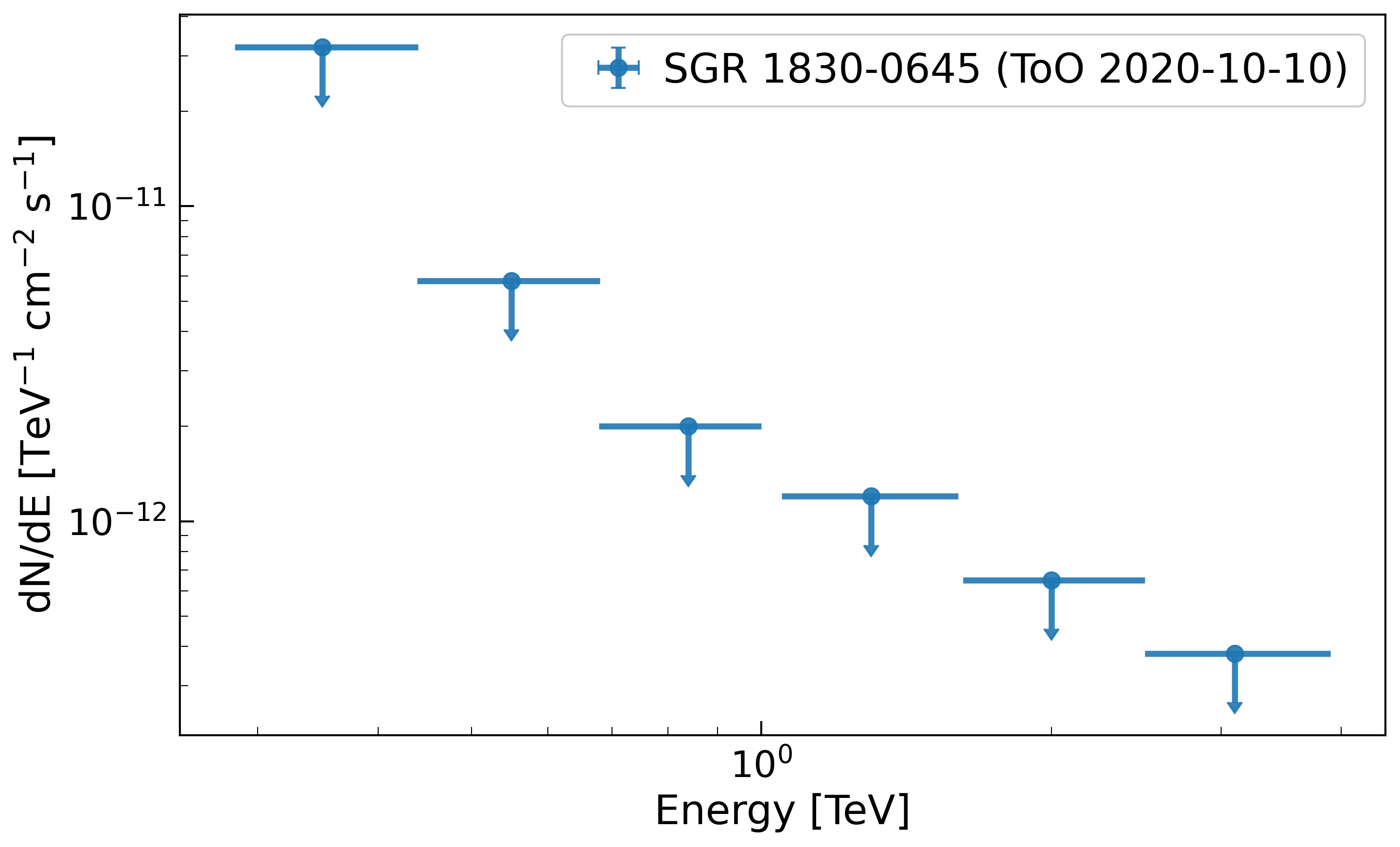}
    \caption{Differential 99\% C.L. upper limits derived from the \hess{} observational data taken during the \fede{ToO observation} of {SGR~1830$-$0645}.}
    \label{fig:sgr1830_diffULs}
\end{figure}

\section{Multiwavelength campaigns}
\label{sec:campaigns}
The second strategy employed by \hess{} to search for FRB sources involves participating in coordinated \mmw{}  campaigns {where  the focus} is on observing known FRB sources, whether repeating or not, to study their host environments and {on} capturing FRBs simultaneously with radio and X-ray facilities. 
By collecting data across different wavelengths simultaneously, {the campaigns aim} to uncover any recurring patterns or {behaviours} associated with the FRB sources.  
{\hess{} has been involved in four \mmw{} campaigns since 2019: Deeper Wider Faster (DWF), two with MeerKAT, and one with INTEGRAL.}

The latter was led by the INTEGRAL {gamma ray instrument} on September 24 -- 28, 2017, {but also} involved the participation of radio (Effelsberg, Nancay, Arecibo, GBT) and optical observatories. {\hess{} conducted observations on the nights of the 26th and 27th of September 2017, contemporaneous with the MAGIC telescopes~\cite{magic}.} However, no FRB was detected by any instrument; therefore, the data collected by \hess{} will not be presented in the following.

\subsection{\fede{Deeper Wider Faster} Campaign} 
\fabian{The Deeper Wider Faster (DWF) campaigns are coordinated \mmw{} and multi-messenger monitoring and follow-up observations of fast transients with a special interest in FRBs~\citep{2019IAUS..339..135A}. They involve a large range of ground and space-based facilities conducting coordinated, simultaneous observations of pre-defined fields. In addition, multiple facilities are on stand-by allowing for fast reaction follow-up in case of interesting detections during these observations. H.E.S.S. participated in both aspects of the DWF campaign discussed here. }
\hess{} participated in a DWF observation campaign from June 23 until June 29, {2019}. \fede{Several facilities observed three pre-defined targets -- Field~1, Field~2, and Field~3 -- centered at (16h26m00s, $-73^{\circ}00^{\prime}00^{\prime\prime}$), (19h08m00s, $64^{\circ}30^{\prime}00^{\prime\prime}$), and (21h44m25s, $-40^{\circ}54^{\prime}00^{\prime\prime}$), respectively. The observations were aimed at detecting potential FRBs and their counterparts. Fields 1 and 2 included the galaxies NGC~6101 and NGC~6744 within their FoV, while Field~3 was centered on the host galaxy of FRB~20180924B~\citep{doi:10.1126/science.aaw5903}.} \ha{This campaign was conducted over 5 nights.} These observations were taken under moderate moonlight {conditions} and served the commissioning of \hess{} moonlight observations. {The H.E.S.S. observation summary, given in \autoref{tab:dwf}{,} serves to search for VHE emission from any detected FRB \ha{or FRB sources}. No FRBs were detected in this campaign.}



\begin{table*}
\centering
\resizebox{0.9\textwidth}{!}{
\begin{tabular}{cccccccc}
\toprule
\toprule
\sc Night & \sc Field & \sc Observation Start & \sc Observation End  \\
\midrule
1 & \fede{Field 1} & 2019-06-23 23:00 & 2019-06-24 00:30 \\ 
2 & \fede{Field 1} & 2019-06-24 23:00 & 2019-06-25 00:30 \\
3 & \fede{Field 1} & 2019-06-25 23:00 & 2019-06-26 01:30 \\
4 & \fede{Field 1} & 2019-06-26 23:00 & 2019-06-27 01:30 \\
5 & \fede{Field 1} & 2019-06-27 23:00 & 2019-06-28 01:30 \\
\midrule
2 & \fede{Field 2} & 2019-06-25 00:30 & 2019-06-25 03:30 \\
3 & \fede{Field 2} & 2019-06-26 01:30 & 2019-06-26 03:30 \\
4 & \fede{Field 2} & 2019-06-27 01:30 & 2019-06-27 03:30 \\
5 & \fede{Field 2} & 2019-06-28 01:30 & 2019-06-28 03:30 \\
\midrule
2 & \fede{Field 3} & 2019-06-25 03:30 & 2019-06-25 04:30 \\
3 & \fede{Field 3} & 2019-06-26 03:30 & 2019-06-26 04:30 \\
4 & \fede{Field 3} & 2019-06-27 03:30 & 2019-06-27 04:30 \\
5 & \fede{Field 3} & 2019-06-28 03:30 & 2019-06-28 04:30 \\
\bottomrule\bottomrule

\end{tabular}}
\caption{{Catalogue} of observations conducted throughout {the 5 H.E.S.S. observation nights of the} 2019 DWF campaign. All observation times are presented in {UTC}.}
\label{tab:dwf}
\end{table*}

{After passing the quality checks, the data on the three fields were analyzed, each night separately. No significant VHE {gamma-ray} signal {was} found for individual nights, nor when stacking all five. The 99\%~C.L. upper limits derived from this analysis are shown in \autoref{fig:DWF_uls}.}


\begin{figure}
    \centering
    \includegraphics[width=0.9\textwidth]{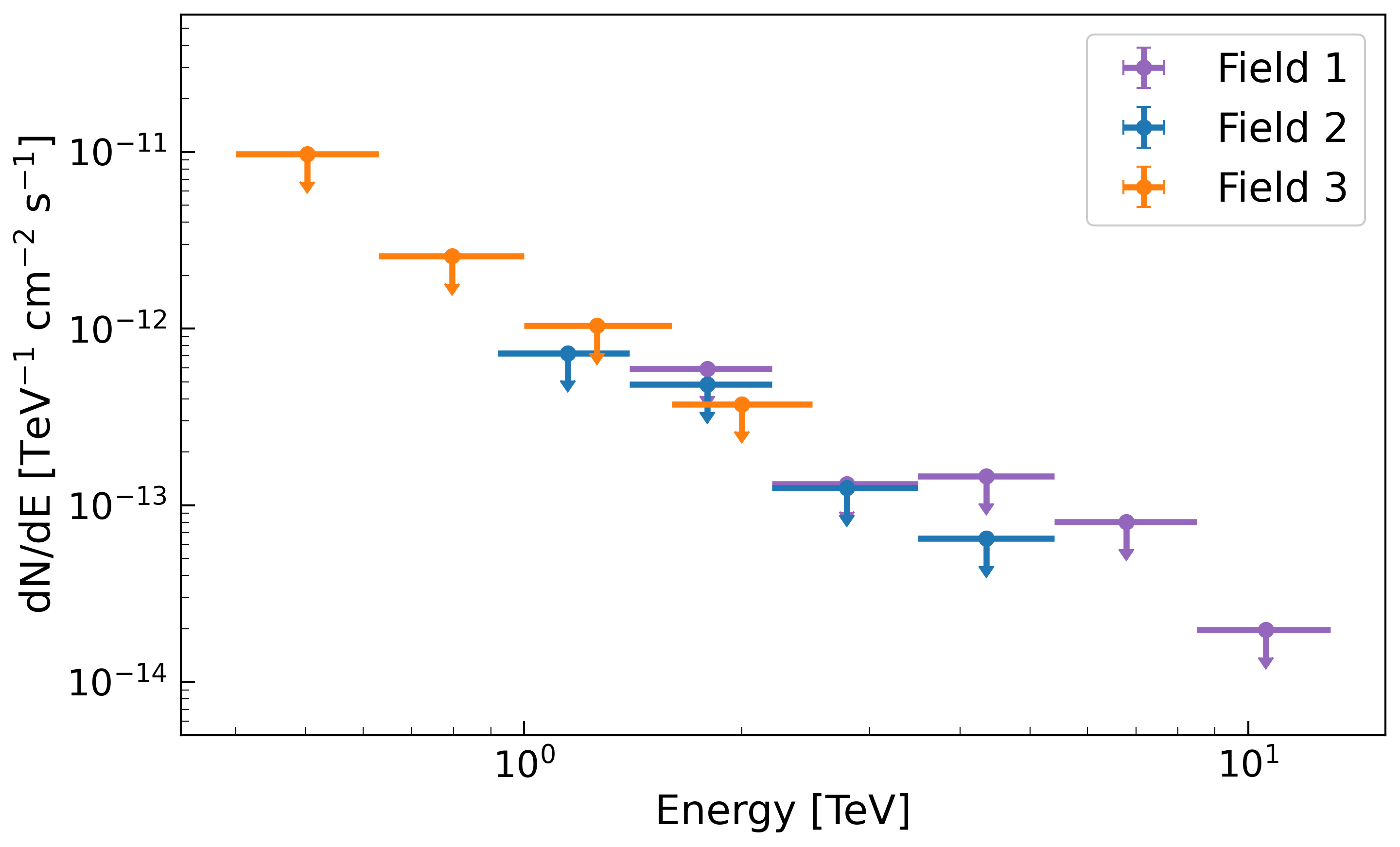}
    \caption{Differential 99\% C.L. upper limits derived from the \hess{} observational data taken on \fede{Field 1, 2, and 3} during the DWF campaign. \fede{The timescales used to derive these limits correspond to the H.E.S.S. observation durations for each source, as specified in the main text and in \autoref{tab:dwf}.}}
    \label{fig:DWF_uls}
\end{figure}

\subsection{MeerKAT Campaigns} 

{\hess{} has participated} in FRB campaigns led by the MeerKAT radio telescope since 2019. MeerKAT, a precursor of the Square Kilometer Array (SKA)~\cite{ska}, is a radio interferometer in the Karoo region of South Africa~\cite{meerkat}. 
 

In 2019, \hess{} joined the first night of campaigns targeting the {southern} repeating FRB~20171019A \ha{(22h17m31.92s, -8d39'32.4')}, conducted over two nights of observations on September 27 and October 18. \ha{The campaign aimed to search for variability through simultaneous observations with H.E.S.S. and MeerKAT, ensuring contemporaneous data in the radio and VHE gamma-ray domains. Additionally, 2~ks of Swift ToO observations were conducted on September 28 (18:37:02–21:52:54 UTC) and October 18 (18:03:00–20:03:00 UTC), providing complementary data in the X-ray and UV ranges.} The analysis of {1.8~hours of }\hess{} data did not reveal any significant excess of gamma-ray emission above the expected background from the direction of FRB~20171019A. {The \fede{flux integral upper limits} were calculated with a 95~\% C.L. and assuming an energy dependence following $E^{-2}$. This resulted in $\Phi_\gamma(E>120~\rm{GeV})<2.10\times 10^{-12}$~cm$^{-2}$s$^{-1}$~\cite{frb171019}.} {Additionally, no emission in radio, X-ray, or ultraviolet wavelengths was detected from the region of FRB20171019A, and upper limits were established~\cite{frb171019}.}

\begin{table*}
\centering
\resizebox{1\textwidth}{!}{
    \begin{tabular}{ccccc}
        \toprule
        \toprule
        \sc Field & \sc MeerKAT Observations & \sc \hess{} Observations &  \sc H.E.S.S. Target & \sc Epoch \\
        \midrule
        \fede{FRB~20200430A} &  28 Jul 2021  18:11 - 20:50 & 28 Jul 2021 18:16 - 20:11 & (15h18m50s, 12d22$^{\prime}$34$^{\prime\prime}$) & 2A\\
        \fede{FRB~20190608B} & 2 Sep 2021 21:00 -- 23:30 & 2 Sep 2021 21:00 -- 23:25 & (22h16m05s, -7d53$^{\prime}$54$^{\prime\prime}$) & 2B\\
        \fede{FRB~20181112A} & 3 Sep 2021 20:45 -- 23:15 & 3 Sep 2021 21:16 -- 22:13 & (21h49m24s, -52d58$^{\prime}$15$^{\prime\prime}$) & 2B\\
        \fede{FRB~20190102C} & 4 Sep 2021 20:15 -- 22:45 & 4 Sep 2021 20:18 -- 22:43 & (21h29m40s, -79d28$^{\prime}$31$^{\prime\prime}$) & 2B\\
        \fede{FRB~20190611B} & 5 Sep 2021 20:15 -- 22:45 & 5 Sep 2021 20:18 -- 22:45 & (21h22m58s, -79d23$^{\prime}$51$^{\prime\prime}$) & 2B\\
        \fede{FRB~20191228A} & 6 Sep 2021 21:30 -- 00:00 & 6 Sep 2021 21:16 -- 00:11 & (22h57m43s, -29d35$^{\prime}$38$^{\prime\prime}$) & 2B\\
        \bottomrule
        \bottomrule
    \end{tabular}}

    \caption{Details of the FRB fields observed during the 2021 MeerKAT campaign. All observation times are presented in \fede{UTC} time.}
    \label{tab:meerkat2021}
\end{table*}

{Subsequently, in 2021, \hess{} participated in a new campaign to observe seven FRB sources in collaboration with MeerKAT, with {X-rays} observations performed by Swift, and optical observations {by} ATOM and MeerLicht (see \autoref{tab:meerkat2021}).}
\fede{The campaign aimed to detect gamma-ray emission or set limits on  the persistent emission.}
The campaign was divided into three epochs as follows: a first epoch with observations carried out solely by MeerKAT, and two epochs (labeled 2A and 2B), with simultaneous observations by MeerKAT, H.E.S.S., and Swift. \fede{Each MeerKAT observation was accompanied by 2~ks of Swift observations.} Epoch 2A observations specifically targeted \fede{FRB~20200430A} and were conducted on July 28, 2021 for 2 hours, while Epoch 2B targeted five different sources (\fede{FRB~20190608B}, \fede{FRB~20181112A}, \fede{FRB~20190102C}, \fede{FRB~20190611B}, and \fede{FRB~20191228A}) for 2.5 hours each night from \ha{September 2-6, 2021.} 
\fede{The epoch 2B campaign prioritized well-localized sources and incorporated a mix of repeating and non-repeating FRB sources.} No significant {gamma-ray emission was detected.} The gamma-ray differential upper limits for the six fields observed by \hess{} are depicted in \autoref{fig:meerkat_uls}. 
{No persistent radio emission was detected by MeerKAT either. While observing the position of {\fede{FRB~20190611B}}, the FRB~20210410D was discovered~\cite{frb210410d}.}

\begin{figure}
    \centering
    \includegraphics[width=0.9\textwidth]{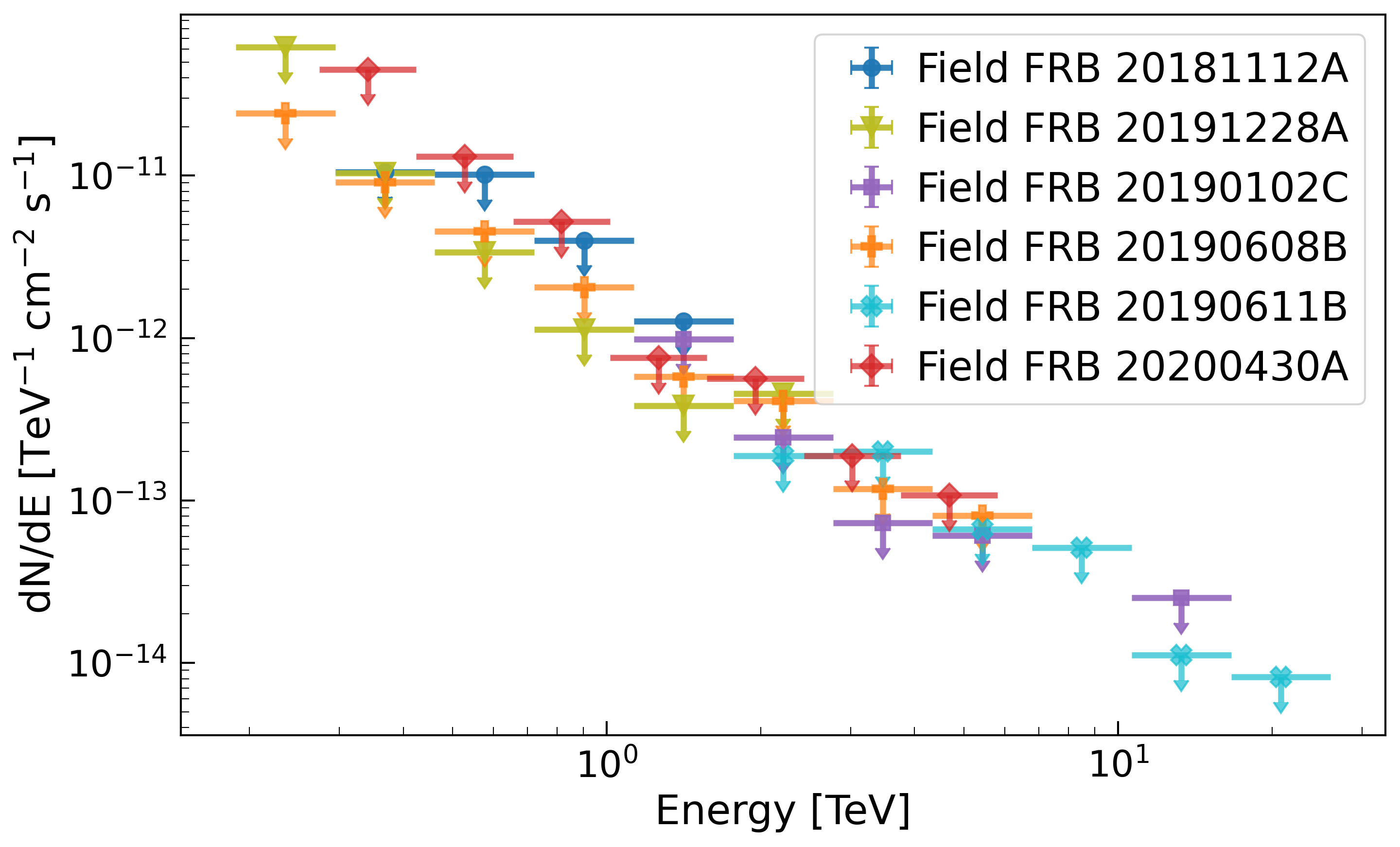}
    \caption{Differential 99\% C.L. upper limits derived from the \hess{} observational data taken on the targets of the MeerKAT campaign of 2019. \fede{The timescales used to derive these limits correspond to the H.E.S.S. observation durations for each source, as specified in the main text.}}
    \label{fig:meerkat_uls}
\end{figure}



\section{Discussion and conclusions}

The H.E.S.S. collaboration has conducted extensive observations targeting localized FRB sources, potential repeating FRBs sources, and \ha{magnetars}. These efforts, which included follow-up observations triggered by radio and X-ray detections and coordinated multi-wavelength campaigns, aimed {at enhancing} our understanding of the environments and potential {emission} mechanisms behind FRBs. However, no significant VHE gamma-ray emission has been detected {from any of the targets}. Nonetheless, the H.E.S.S. collaboration's FRB {programme} remains crucial for investigating these intriguing astrophysical sources~\cite{2019ARA&A..57..417C, 2019A&ARv..27....4P} and shedding light on their still unknown origins and emission mechanisms~\cite{2020Natur.587...45Z, 2021SCPMA..6449501X, 2021Univ....7...56L, 2023RvMP...95c5005Z}. The H.E.S.S. {programme} complements the one performed by VERITAS~\cite{2019ICRC...36..698H, 2021arXiv210809354L}. VERITAS's proximity to the radio telescope CHIME, only separated by $8.7^{\circ}$ in longitude, allows it to observe sources approximately 30~minutes past culmination as they transit through the CHIME field of view. 

Among the fields observed by H.E.S.S. from 2015 to 2022, \fede{nine} \ha{contain} localized FRBs, for which information on the radio burst, the host galaxy and chance association is provided in the literature and in Table~\ref{tab:localized_frbs}. The {radio} luminosity of the extragalactic \fede{FRB bursts} mentioned in this work is between $10^{41}$~erg~s$^{-1}$ and $10^{45}$~erg~s$^{-1}$. We summarize the information on these \fede{nine} localized FRBs in \autoref{tab:localized_frbs}. Using H.E.S.S. observations, we can set a limit on the \ha{continuous VHE emission of the environment} associated with these sources. We calculate the gamma-ray luminosity {upper limits} by correcting the 99\% C.L. integral flux upper-limits for {EBL} absorption using the redshift values in \autoref{tab:localized_frbs}. The H.E.S.S. upper limit on the VHE luminosity of {extragalactic FRB positions} ranges between $10^{44}$~erg s$^{-1}$  and $10^{48}$~erg s$^{-1}$ for \ha{redshifts} raging between 0.11 and 0.49. {The upper limits are comparable to the upper limits derived by Fermi-LAT~\cite{2023A&A...675A..99P}}.  On the other hand,~\cite{10.1093/mnras/stw1328} expected continuous luminosities can reach $10^{39}$~erg~s$^{-1}$ at 1 TeV, lying several orders of magnitude below the H.E.S.S. upper limits derived in this work. \ha{ranging beetween  $10^{44}$~erg s$^{-1}$  and $10^{48}$~erg s$^{-1}$. This continuous luminosity can reach $10^{42}$~erg~s$^{-1}$ at 1 GeV requiring a lower energy threshold and a better sensitivity that might be achievable with future instruments. In comparison observation on FRB 20121102 and FRB 20220912A place upper limits of $3 \times 10^{41}$~erg~s$^{-1}$ ~\cite{2017ApJ...846...80S} and $1.4 \times 10^{41}$~erg~s$^{-1}$~\cite{2024ApJ...974..170C} in the X-ray domain respectively.}

As for the transient VHE emission expected from the FRB source,~\cite{Lyubarsky:2014jta}, expects a total energy of around $10^{48}$~erg for VHE bursts \ha{accompanying} FRBs  {in the TeV domain}. This emission is expected to be detectable by Cherenkov Telescopes from a distance of 100 Mpc. \ha{From~\cite{HESS_SGR1935} the H.E.S.S. sensitivity to such burst is $10^{-9}$~erg cm$^{-2}$, requiring a VHE total emission above $10^{46}$~erg to $10^{48}$~erg from the sources in this sample to be able to be detected by H.E.S.S.} \ha{In comparison, observation of FRB 20200120E with Konus-Wind place upper bounds of $2 \times 10^{44}$~erg in the 20 - 1500 keV energy range. Moreover, observation on FRB 20180916B in the optical and X-ray ranges suggest respective ratios of  $E_{optical}/E_{radio} < 10^2$ and $E_{X}/E_{radio} < 10^7$~\cite{2023A&A...676A..17T}.}  Therefore, we conclude that probing VHE emission from FRB sources with {\hess{}} {provides valuable constraints on theoretical models for these enigmatic bursts, {especially for multi-wavelength campaigns where the FRB might occur during VHE observation time.}} \ha{\hess{} remains the only instrument located in the Southern Hemisphere that can probe possible VHE emission from transient events making it invaluable for the worldwide effort to search for FRB counterparts.}

{Moreover, the application of new tools to search for transient phenomena in gamma-ray data~\cite{galaxies11040088} and the advent of wide-field radio surveys will enable a more complete picture of FRBs. A significant increase in the number of FRB events will be facilitated by {SKA} and the new and expanded interferometer CHIME/CHORD~\cite{chime}, which will expand the samples of events with accurate localization and will improve the characterization of their radio properties as well as the investigation of their local environments~\cite{2020MNRAS.497.4107H}.} {In the future, the commissioning of the Cherenkov Telescope Array Observatory (CTAO) will further enhance the likelihood of detecting emission from FRB sources in the VHE domain.} Compared with current VHE instruments, CTAO will be an order of magnitude more sensitive above 1 TeV, {with} several times better sensitivity in the range 30--100~GeV where the effects of EBL absorption are lower. With better sensitivity, CTAO will be able to resolve shorter time-scale variability by collecting more events and the hunt for FRB counterparts becomes much more efficient with a luminosity upper limit that can go down to $10^{42}$ -- $10^{43}$~erg s$^{-1}$ for \fede{FRB~20190608B}, the closest case from this sample.

\begin{landscape}
\begin{table*}
\centering 
\begin{tabular}{ccccccccc}
\toprule
\toprule

\multirow{2}{*}{\sc Source} & \multirow{2}{*}{\sc Detection}  & \multirow{2}{*}{\sc{Repeater}} & \multirow{2}{*}{\sc Redshift} & \sc $L_{\rm{r, FRB}}$  &\sc $L_{\rm{r, host}}$ &  \ha{Chance } & \multirow{2}{*}{\sc Ref.} & \multirow{2}{*}\sc $L_{\rm{VHE, host}}$  \\

&   &  & & [erg s$^{-1}$] & [erg s$^{-1}$] & \ha{Association} & & [erg s$^{-1}$]\\

\midrule

\fede{FRB~20190806A} & UTMOST &  No & $\lesssim 0.32$ & -- & -- & -- & \citep{2019ATel12995....1G} & $<3.61 \times 10^{45}$  \\ 

\fede{FRB~20150418A} & PARKES & No &  $0.4920$ & $<1.3 \times 10^{42}$ & $5.6 \times 10^{39~*}$ &  0.0018 & \citep{2016Natur.530..453K, 2016AA...593L..16G, 2020ApJ...903..152H}  & $<4.35 \times 10^{46}$  \\ 


\fede{FRB~20180924B} & ASKAP & No   & 0.3212  & $3.27 \times 10^{43}$ & $< 3.49\times 10^{38} $  & 0.0018 & 
\citep{frbs_fermi, 2020ApJ...895L..37B, 2020ApJ...903..152H}& $<2.41 \times 10^{45}$   \\ 


\fede{FRB~20200430A} & ASKAP & No &  0.1600 & -- & $< 5.01 \times 10^{37}$ & 0.0051 & \citep{frbs_fermi, 2020ApJ...895L..37B, 2020ApJ...903..152H} & $<3.84 \times 10^{44}$  \\ 

\fede{FRB~20190608B} & ASKAP & No & 0.1178 & $9.88 \times 10^{41}$ & $< 2.31 \times 10^{37}$ & 0.0016& \citep{frbs_fermi, 2020ApJ...895L..37B, 2020ApJ...903..152H} & $<1.47 \times 10^{44}$ \\

\fede{FRB~20181112A} & ASKAP & No &  0.4755 & $2.69 \times 10^{45}$ & $<8.37\times 10^{38}$& 0.0257 & \citep{frbs_fermi, 2020ApJ...895L..37B, 2020ApJ...903..152H} & $<1.53 \times 10^{46}$ \\ %

\fede{FRB~20190102C} & ASKAP & No &   0.2912 & $2.0 \times 10^{44}$  & $<2.71\times 10^{38}$& 0.0050 & \cite{frbs_fermi, 2020ApJ...895L..37B, 2020ApJ...903..152H} & $<2.82 \times 10^{46}$  \\ %

\fede{FRB~20190611B} & ASKAP & No &   0.3778 & $5.03 \times 10^{43}$ & -- & 0.0169 & \cite{frbs_fermi, 2020ApJ...903..152H}  & $<5.89 \times 10^{47}$  \\ %

\fede{FRB~20191228A} & ASKAP & No &  0.2432 & $1.49 \times 10^{44}$ & $<2.21 \times 10^{38~*}$ & 0 & \cite{FRB20191228}  & $<1.07 \times 10^{45}$  \\


\ha{FRB~20200428D} & \multirow{2}{*}{CHIME} & \multirow{2}{*}{No} &  \multirow{2}{*}{6.6–12.5~kpc} & \multirow{2}{*}{$3\times10^{36}$} & \multirow{2}{*}{--} & \multirow{2}{*}{--} & \multirow{2}{*}{\cite{HESS_SGR1935}}& \multirow{2}{*}{$<1.27 \times 10^{34}$} \\
(SGR~1935+2154) &  &  &  &  & &  &  \\ 

\bottomrule
\bottomrule
\end{tabular}
\\
\vspace{2mm}

$^{*}$ \fede{{Steady} radio luminosity from resolved region inside the host.}
\caption{{Host galaxy properties for localized FRBs studied by \hess{}} \fede{Spectroscopic redshifts are reported to four significant digits (typical
uncertainty)~\cite{2020ApJ...903..152H}.}\ha{The chance association of the FRB to their host galaxies are taken from the works cited in Ref. $L_{\rm{VHE,host}}$ constrains the VHE gamma-ray luminosity from the region of the FRB during the H.E.S.S. observations and was obtained assuming a E$^{-2}$ spectrum and after correcting for EBL absorption.}}
\label{tab:localized_frbs}
\end{table*}

\end{landscape}

\acknowledgments

The support of the Namibian authorities and of the University of
Namibia in facilitating the construction and operation of H.E.S.S.
is gratefully acknowledged, as is the support by the German
Ministry for Education and Research (BMBF), the Max Planck Society,
the Helmholtz Association, the French Ministry of
Higher Education, Research and Innovation, the Centre National de
la Recherche Scientifique (CNRS/IN2P3 and CNRS/INSU), the
Commissariat à l’énergie atomique et aux énergies alternatives
(CEA), the U.K. Science and Technology Facilities Council (STFC),
the Irish Research Council (IRC) and the Science Foundation Ireland
(SFI), the Polish Ministry of Education and Science, agreement no.
2021/WK/06, the South African Department of Science and Innovation and
National Research Foundation, the University of Namibia, the National
Commission on Research, Science \& Technology of Namibia (NCRST),
the Austrian Federal Ministry of Education, Science and Research
and the Austrian Science Fund (FWF), the Australian Research
Council (ARC), the Japan Society for the Promotion of Science, the
University of Amsterdam and the Science Committee of Armenia grant
21AG-1C085. We appreciate the excellent work of the technical
support staff in Berlin, Zeuthen, Heidelberg, Palaiseau, Paris,
Saclay, Tübingen and in Namibia in the construction and operation
of the equipment. This work benefited from services provided by the
H.E.S.S. Virtual Organisation, supported by the national resource
providers of the EGI Federation.




\bibliographystyle{JHEP}
\bibliography{FRBreferences}





\end{document}